\documentclass[10pt,onecolumn,preprintnumbers,amsmath,amssymb]{revtex4}
\usepackage{graphicx}
\usepackage{color}

\begin{document}
\title{Generalized uncertainty principle and stochastic gravitational wave background spectrum}
\author{Mohamed Moussa}
\email{mohamed.ibrahim@fsc.bu.edu.eg}
\affiliation{Physics Department, Faculty of Science, Benha University, Benha 13518, Egypt}
\author{Homa Shababi}
\email{h.shababi@scu.edu.cn}
\affiliation{Center for Theoretical Physics, College of Physical Science and Technology,
Sichuan University, Chengdu 610065, P. R. China}
\author{Ahmed Farag Ali $^1$}
\email[email: ]{ahmed.ali@fsc.bu.edu.eg}
 \affiliation{$^{4}$Physics Department,  Faculty  of  Science, Benha  University,  Benha,  13518,  Egypt}

\begin{abstract}

This paper concerned with the effect of generalized uncertainty principle (GUP) on the stochastic gravitational wave (SGW) background signal that produced during first order cosmological QCD phase transition in early universe. A modified formula of entropy is used to calculate the temporal evolution of temperature of the universe as a function of the Hubble parameter.
The pressure that results from the recent lattice calculations, which provides parameterizations of the pressure due to $u,~d,~s$ quarks and gluons, with  trace anomaly is used to describe the equation of state around QCD epoch.
A redshift in the peak frequency of SGW at current epoch is calculated. The results indicate an increase in the frequency peak due to GUP effect, which improves the ability to detect it.
Taking into account bubble wall collisions (BWC) and turbulent magnetohydrodynamics (MHD) as a source of SGW, a fractional energy density is investigated.
It is found that the GUP effect weakens the SGW signal generated during QCD phase transition in comparison to its counterpart in the absence of GUP.
These results support understanding the cosmological QCD phase transition and test the effectiveness of the GUP theory.

\end{abstract}

\maketitle

\section{introduction}
One of the most important physical phenomena that have attracted both attention in theoretical physicists and observational astrophysicists is the discovery of gravitational waves (GWs).
It is known that, from the merger of black holes by Laser Interferometer Gravitational-Wave Observatory (LIGO) collaboration, GWs usher in a new era in astronomy and cosmology \cite{LIGO}. LIGO detectors are structured in such a way that they operate in high frequency range ($10-10^{3}$ HZ), for detecting sources like compact binary inspirals. Also, to identify lower frequency sources ($10^{-5}-1$ HZ), GWs signals from sources such as Supernovae and eLISA experiment \cite{Klein}, have been proposed. Moreover, some setups including Square Kilometer Array (SKA) \cite{Kramer,Dewdney} and Pulsar Timing Array (PTA) \cite{IPTA} can measure frequency range even lower than $10^{-5}$ Hz (around $10^{-9}$ Hz). These divisions of experiments indicate that although the strongest gravitational waves are generated by events such as colliding black holes, stellar core collapses (supernovae), coalescing neutron stars and etc, there also be a random background of GWs, which is called stochastic gravitational wave  (SGW) background, and characterized by sharp frequency component.
Detectors of multi-wavelength can assist observation those waves, and the results may play a major role in a deeper understanding of the universe.

Although low frequencies are difficult to be observed experimentally, modeling their sources is theoretically very important because they can provide vital information about the early stages of the universe \cite{Anand}.
In this regard, we can refer to the SGW background that contains information about the early universe. It is agreed that long-duration first-order cosmological phase transitions could be a potential source of very low-frequency gravitational wave background \cite{Witten,Hogan,Hogan2,Turner}. According to standard particle physics, there are at least two phase transitions, the electroweak phase transition at $T\sim 100$ GeV, which is accompanied by breaking the electroweak symmetry, and the QCD phase transition at $T\sim 0.1$ GeV, which breaks the chiral symmetry. Although it is known that these transitions are done through a smooth crossover, there are models beyond the standard model that can produce strong first-order transition at the electroweak and QCD scales \cite{10,11,12,13,14,15,16,17,18,19,20}. So far no agreement has been reached on the exact critical temperature of these transitions or their nature. The main goal in the study of the hadronization process, relevant to strong QCD phase transition, is to reach the equation of state (EoS) governing two different phases, quark-gluon plasma (QGP) and hadronic gas (HG). According to the recent results of QCD lattice, in the presence of strong interactions, the pressure $P$ is no longer equal to its value in the radiation dominated epoch \cite{DeTar}. This modification is specified in terms of trace anomaly which can lead to some exciting cosmic results such as the prediction of Weakly Interacting Massive Particles (WIMPs), and pure glue lattice QCD calculations \cite{Drees}.

Recently, some research has been guided in this direction. The authors in \cite{bc1}, studies the effect of equation of state of relativistic particles along with QCD equation of state, that emerged by parameterization of the pressure due to $u,~d,~s$ quarks and gluons with the energy density that computed from trace anomaly. It is found that the rate of expansion of the universe is decreased, the gravitational wave signal is increased by almost $\textit{50}$ percent and the peak frequency redshift to current time is changed by $\textit{25}$ percent. In \cite{bc2}, a fractional energy density and a peak frequency redshift at current time of SGW is investigated using effective QCD equation of state of three chiral quark flavors $u,~d,~s$ including chemical potential and finite temperature. It is obtained that as the chemical potential increases, the frequency and amplitude of SGW signal that received today, were increased. The effective contribution of the chemical potential of quarks to the wave detection is also discussed.

In this paper, we investigate the effect of GUP on the SGW background signal which is emanated from the first order cosmological QCD phase transition in early universe. The entropy of photon gas that modified by the GUP is used in the calculation of temporal evolution of the temperature of the universe as a function of the Hubble parameter. Hubble parameter and the peak frequency of the SGW frequency red shifted with its corresponding value at current time are calculated. Within GUP framework and taking into account bubble wall collisions and turbulent magnetohydrodynamics in plasma as a source of SGW at the epoch of phase transition, a fractional energy density of SGW are investigated. It is found that the GUP effect reduces the SGW signal in comparison to its counterpart in the absence of GUP. These results can shed light on a better understanding of the cosmological QCD phase transition and show the importance of the GUP theory.

\section{GUP and photons gas}
A different approaches to quantum gravity such as string theory, non-commutative geometry and black hole physics predict existence of a minimal measurable length. This in turn will produce an essential modification of the Heisenberg uncertainty principle to what so called generalized uncertainty principle (GUP) \cite{d1,d2,d3,d4,d5,d6}. According to this approach, it is suggested that \cite{d7}
\begin{equation}\label{1}
\Delta x_i \Delta p_j\geq\frac{\hbar }{2}{\delta }_{ij}(1+\beta
[\left(\Delta p\right)^{2}+ \langle p\rangle^2]~),
\end{equation}
where $\beta$ is the GUP parameter and defined as $\beta=\frac{\beta_{0}}{m_p^2c^2}$ which $m_p$ is the Planck mass and
$\beta_{0}$ is of the order of unity. Then, the absolute minimal measurable length is obtained by saturating the
above inequality as $(\Delta x)_{min}=\hbar\sqrt\beta\sqrt{1+\beta \langle p\rangle^2}$ and so
$(\Delta x)_{min}=\hbar\sqrt\beta=\sqrt{\beta_0}\ell_p$ for $\langle
p\rangle=0$, where $\ell_p$ is the Planck length. It should be noted that $(\Delta x)_{min}$ is not exactly equal to the Planck
length but it is at the same order, depending on whether $\beta_0$ is bigger or smaller than one. It is known that, the exact value of $\beta_0$ is not precisely known but it is extensively investigated and expected to be obtained in future experiments. The uncertainty relation (\ref{1}) leads to the following deformed commutation relation, namely
\begin{equation}\label{2}
\left [x_{i},p_{j}\right]=i\hbar\delta_{ij}\left(1+\beta
p^{2}\right),
\end{equation}
where $p^2=\sum_i p_i^2$. Also, from the Jacobi identity, this equation ensures $\left [x_{i},x_{j}\right]=0$ and $\left [p_{i},p_{j}\right]=0$. Then, as a result of Eq. (\ref{2}), GUP modifies
the physical momentum as
\begin{equation}\label{3}
p_{i}=p_{i0}\left(1+\beta {p_{i0}}^2\right),~~~~~ x_i=x_{i0}
\end{equation}
while $x_{i0}$ and $p_{i0}$ satisfy the usual canonical commutation relations $[x_{i0}, p_{j0}] =i\hbar\delta_{ij}$. Thus we can consider $p_i$ is the momentum in Planck scale and $p_{i0}$ is the momentum at low scale. On the other hand, according to the Liouville theorem, the number of quantum states inside phase space should not be changed with time evolution within GUP framework. Thus GUP should adjust the density of state to agree with Liouville theorem, which will have an effect on the thermodynamics properties of quantum systems. This way, the number of quantum states should be modified, such that \cite{d8}
\begin{equation}
\frac{V}{(2\pi)^3} \int_0^{\infty}d^3p \rightarrow \frac{V}{(2\pi)^3} \int_0^{\infty}\frac{d^3p}{(1+\beta p^2)^3}.
\end{equation}
Within framework of mentioned GUP,  thermodynamic properties of photons gas will be investigated. Then we can get the modified entropy of photons due to GUP effect, which is the aim of this section. According to above modified phase space, the grand canonical partition function of photons can be written as
\begin{eqnarray}
\nonumber \ln{Z} =-\frac{Vg_{\pi}}{2\pi^2}\int_0^{\infty}\ln\left[1-e^{-\frac{p}{T}}\right]\frac{p^2dp}{(1+\beta p^2)^3},~~~~ \\
\simeq-\frac{Vg_{\pi}}{2\pi^2}\int_0^{\infty}\ln\left[1-e^{-\frac{p}{T}}\right](1-3\beta p^2)~p^2dp,
\end{eqnarray}
where $g_{\pi}$ is the number of degrees of freedom. After some manipulation, we obtain
\begin{eqnarray}
\nonumber \ln{Z} =\frac{Vg_{\pi}}{2\pi^2}\frac{1}{T}\int_0^{\infty}\frac{1}{e^{\frac{p}{T}}-1}~\left[\frac{1}{3}p^3-\frac{2}{5}\beta p^5\right]~dp \\
=\frac{1}{90}Vg_{\pi}\pi^2T^3- \frac{4}{105}\beta Vg_{\pi}\pi^4T^5.~~~~~~~~~~~
\end{eqnarray}
Using the modified partition function, the entropy of photon gas can be written as
\begin{equation}
S=\frac{\partial}{\partial T}(T\ln{Z})=\frac{2}{45}g_{\pi}\pi^2T^4- \frac{8}{35}\beta g_{\pi}\pi^4T^6.
\end{equation}
In a good approximation, the entropy expression can be rewritten as
\begin{equation}
S=\frac{g_{\pi}\frac{2\pi^2}{45}T^3}{1+\frac{37}{7}\pi^2\beta T^2}.
\end{equation}
In the limit of $\beta\rightarrow0$, we can recover standard form $S=g_{\pi}\frac{2\pi^2}{45}T^3$.

\section{Modified stochastic gravitational wave spectrum}

It is agreed that the SGW is generated at the epoch of QCD and electroweak phase transition and propagate to the current epoch. Here we interested in SGW that produced at the first order QCD phase transition. To investigate the observable spectrum of SGW, we will use the assumption that the universe has expanded adiabatically which results in the constancy of the entropy per comoving volume, such that $\frac{\dot{S}}{S}=0$. We will use the entropy of photons gas where the photons dominate universe more than baryons. Thus, the entropy can be estimated using the entropy of QGP state in the previous section, i.e.
\begin{equation}\label{b1}
S\sim a^3 \frac{g_s\frac{2\pi^2}{45}T^3}{1+\frac{37}{7}\pi^2\beta T^2},
\end{equation}
where $g_s$ is the effective number of relativistic degrees of freedom involved in the entropy density and $a$ is the scale factor. Employing the adiabatic condition $\frac{\dot{S}}{S}=0$, the formula of time variation of temperature can be obtained
\begin{equation}\label{b2}
\frac{dT}{dt}=-3H\left(\frac{3}{T}+\frac{1}{g_s}\frac{dg_s}{dT}-\frac{72\pi^2\beta T}{7+36\pi^2\beta T^2}\right)^{-1},
\end{equation}
which $H$ is the Hubble parameter. It is clear that the ordinary case is recovered in the limit of $\beta \rightarrow 0$ \cite{bc1}. Also, we can rewrite Eq. (\ref{b2}) in terms of scale factor as
\begin{equation}\label{b3}
\frac{a_*}{a_0}=\exp{\left[\int_{T_*}^{T_0}\frac{1}{T}\left(1+\frac{T}{3g_s}\frac{dg_s}{dT}-\frac{24\pi^2\beta T^2}{7+36\pi^2\beta T^2}\right)~dT\right]},
\end{equation}
where $" * "$ refers to quantities at the epoch of phase transition and $" ~0~ "$ denotes quantities at the current time. Based on the fact that the SGW are ultimately decoupled to the rest of the universe, the energy density of the SGW can satisfy Boltzmann equation $\frac{d}{dt}\left(\rho_{gw}a^4\right)=0$. Applying Eq. (\ref{b3}) and Boltzmann equation, the energy density of SGW at current time reads
\begin{eqnarray}\label{b4}
\nonumber \rho_{gw}(T_0)=\rho_{gw}(T_*) \left(\frac{a_*}{a_0}\right)^4, ~~~~~~~~~~~~~~~~~~~~~~~~~~~~~~~~~~~~~~~~~~~~~~~~~~~~~~~\\
=\rho_{gw}(T_*)~\exp{\left[\int_{T_*}^{T_0}\frac{4}{T}\left(1+\frac{T}{3g_s}\frac{dg_s}{dT}-\frac{24\pi^2\beta T^2}{7+36\pi^2\beta T^2}\right)~dT\right]}.
\end{eqnarray}
Let us define the density parameter of SGW at phase transition epoch and current time as $\Omega_{gw*}=\frac{\rho_{gw}(T_*)}{\rho_{cr}(T_*)}$ and $\Omega_{gw}=\frac{\rho_{gw}(T_0)}{\rho_{cr}(T_0)}$, respectively, where $\rho_{cr}$ is the critical density. So, using Eq. (\ref{b4}), we obtain
\begin{equation}\label{b5}
\Omega_{gw}=\Omega_{gw*}\left(\frac{H_*}{H_0}\right)^2~\exp{\left[\int_{T_*}^{T_0}\frac{4}{T}\left(1+\frac{T}{3g_s}\frac{dg_s}{dT}-\frac{24\pi^2\beta T^2}{7+36\pi^2\beta T^2}\right)~dT\right]},
\end{equation}
where
\begin{equation}\label{b6}
\left(\frac{H_*}{H_0}\right)^2=\frac{\rho_{cr}(T_*)}{\rho_{cr}(T_0)}.
\end{equation}
In order to determine the evaluation of Hubble parameter from phase transition epoch to the current time, we need to use the continuity equation, namely:
\begin{equation}\label{b7}
\dot{\rho}_t=-3H\rho_t\left(1+\frac{P_t}{\rho_t}\right),
\end{equation}
where $\rho_t$ and $P_t$ are the total energy density and total pressure density of the universe at cosmic time $t$, respectively and dot denotes the derivative with respect to cosmic time. Now using Eq. (\ref{b2}), Eq. (\ref{b7}) can be written in terms of temperature as
\begin{equation}\label{b8}
\frac{d\rho_t}{\rho_t}=\frac{3}{T}\left(1+\omega_{eff}\right)\left(1+\frac{T}{3g_s}\frac{dg_s}{dT}-\frac{24\pi^2\beta T^2}{7+36\pi^2\beta T^2}\right)dT,
\end{equation}
where $\omega_{eff}=\frac{P_t}{\rho_t}$ is the effective equation of state parameter which may depend on temperature. The critical energy density of radiation $\rho_{cr}(T_*)$, at the epoch of phase transition can be obtained by integration from early time $T_r$, where the radiation dominated, to the time of phase transition $T_*$, as
\begin{equation}\label{b9}
\rho_{cr}(T_*)=\rho_r(T_r)\exp{\left[\int_{T_r}^{T_*}\frac{3(1+\omega_{eff})}{T}\left(1+\frac{T}{3g_s}\frac{dg_s}{dT}-\frac{24\pi^2\beta T^2}{7+36\pi^2\beta T^2}\right)~dT\right]}.
\end{equation}
Then, putting Eq. (\ref{b9}) into Eq. (\ref{b6}), one can get
\begin{equation}\label{b10}
\left(\frac{H_*}{H_0}\right)^2=\Omega_{r0}\frac{\rho_r(T_r)}{\rho_r(T_0)}\exp{\left[\int_{T_r}^{T_*}\frac{3(1+\omega_{eff})}{T}\left(1+\frac{T}{3g_s}\frac{dg_s}{dT}-\frac{24\pi^2\beta T^2}{7+36\pi^2\beta T^2}\right)~dT\right]},
\end{equation}
where $\Omega_{r0}=\frac{\rho_r(T_0)}{\rho_{cr}(T_0)}$ is the current value of fractional energy density of radiation with the value almost equal $\Omega_{r0}=8.5\times 10^{-5}$. Now, applying Boltzmann equation, it can be proved that $\frac{\rho_r(T_r)}{\rho_r(T_0)}=(\frac{a_0}{a_r})^4$. Hence, Eq. (\ref{b10}) can be rewritten as
\begin{equation}\label{b11}
\left(\frac{H_*}{H_0}\right)^2=\Omega_{r0}\left(\frac{a_0}{a_r}\right)^4\exp{\left[\int_{T_r}^{T_*}\frac{3(1+\omega_{eff})}{T}\left(1+\frac{T}{3g_s}\frac{dg_s}{dT}-\frac{24\pi^2\beta T^2}{7+36\pi^2\beta T^2}\right)~dT\right]}.
\end{equation}
Substituting Eq. (\ref{b11}) into Eq. (\ref{b5}), we get
\begin{eqnarray}\label{b12}
\nonumber \Omega_{gw}=\Omega_{r0} ~\Omega_{gw*}
~\exp{\left[\int_{T_*}^{T_r}\frac{4}{T}\left(1+\frac{T}{3g_s}\frac{dg_s}{dT}-\frac{24\pi^2\beta T^2}{7+36\pi^2\beta T^2}\right)~dT\right]}\times  ~~~\\
~\exp{\left[\int_{T_r}^{T_*}\frac{3(1+\omega_{eff})}{T}\left(1+\frac{T}{3g_s}\frac{dg_s}{dT}-\frac{24\pi^2\beta T^2}{7+36\pi^2\beta T^2}\right)~dT\right]}.
\end{eqnarray}
At this point, we want to determine the functional form of the effective equation of state, $\omega_{eff}$. It is known that for the ultra-relativistic gas with non-interacting particles, $\omega_{eff}=\frac{1}{3}$. In this case and without GUP effect, the above two relations will be
\begin{equation}\label{bb11}
\left(\frac{H_*}{H_0}\right)^2=\Omega_{r0}\left(\frac{T_*}{T_0}\right)^4\left[\frac{g(T_*)}{g(T_0)}\right]^{\frac{4}{3}},~~~~~\omega_{eff}=\frac{1}{3}, ~~~~\beta=0.
\end{equation}
\begin{equation}\label{bb12}
\Omega_{gw}=\Omega_{r0} ~\Omega_{gw*},~~~~~\omega_{eff}=\frac{1}{3}, ~~~~\beta=0.
\end{equation}
However, around $T\sim $ a few hundred MeV, the relation $\omega_{eff}=\frac{1}{3}$ deviates because the effect of QCD interaction comes to play \cite{Anand}. Thus, we will use QCD equation of state that results from the parametrization of the pressure due to the strong interactions between $u,~d,~s$ quarks and gluons, as \cite{b1}
\begin{equation}\label{b13}
\frac{P}{T^4}=F(T)=\frac{1}{2}\left(1+\tanh{[c_{\tau}(\tau-\tau_0)]}\right)
\frac{p_i+\frac{a_n}{\tau}+\frac{b_n}{\tau^2}+\frac{c_n}{\tau^4}}{1+\frac{a_d}{\tau}+\frac{b_d}{\tau^2}+\frac{c_d}{\tau^4}},
\end{equation}
where $\tau=\frac{T}{T_c}$ with $T_c=154~MeV$ is the QCD transition temperature and $p_i=\frac{19\pi^2}{36}$ is the ideal gas value of $\frac{P}{T^4}$ for QCD interaction with three massless quarks. It is worth noting that the function $\tanh{x}$ approaches unity for large values of $x$, therefore Eq. (\ref{b13}) approaches the ideal gas law at $T \gg T_c$, i.e. $\tau \gg 1$.

Ref. \cite{b2} showed that Eq. (\ref{b13}) matches well with available perturbation calculations at higher temperatures. According to this, the values of numerical coefficients in Eq. (\ref{b13}) which describe the contributions from massless $u,~d,~s$ quarks and gluons for all temperatures above $100~MeV$, are $c_{\tau}=3.6706$, $\tau_0=0.9761$, $a_n=-8.7704$, $b_n=3.9200$, $c_n=0.3419$, $a_d=-1.2600$, $b_d=0.8425$ and $c_d=-0.0475$. Then, we can calculate QCD contribution in the effective equation of state, $\omega_{eff}$, at the temperature range above $100~MeV$, using Eq. (\ref{b13}) with the trace of the energy momentum tensor or called the trace anomaly \cite{b2}
\begin{equation}\label{b13b}
\frac{\rho-3P}{T^4}=T\frac{d}{dT}\left(\frac{P}{T^4}\right).
\end{equation}
Thus, the effective equation of state with QCD effect is obtained
\begin{equation}\label{b14}
\omega_{eff}=\frac{P}{\rho}=\left[\frac{T}{F(T)}\frac{dF(T)}{dT}+3\right]^{-1}.
\end{equation}
To investigate the results more, we have plotted the effective equation of state function versus transition temperatures $T_{*}$, in Fig. (1). As it is shown in this figure, the trace anomaly matches ideal gas at $\sim5~GeV$. This behaviour indicates that the effect of trace anomaly can not be ignored near QCD transition epoch and it should be taken into account.
Then in Fig. (2), we have depicted relative Hubble parameter as a function of transition temperature $T_*$, in case of $\beta=0$, with trace anomaly, shown with solid curve, and without trace anomaly, shown with dashed curve where $\frac{H_*}{H_0}\sim T_*^2$. This figure implies that, under the influence of QCD, Hubble parameter changes slower than $T_*^2$ before $\sim 0.13~GeV$ and then changes faster at higher temperature till matches with $T_*^2$-curve at $\sim 5~GeV$. We have used $T_r=10^4~GeV$. It is clear that the effect of QCD appears at low temperatures, so we will use equation of state of trace anomaly in all our calculations, i.e. Eq. (\ref{b14}).
\begin{figure}[!tbp]
  \centering
  \begin{minipage}[b]{0.4\textwidth}
    \includegraphics[width=7cm, height=5cm]{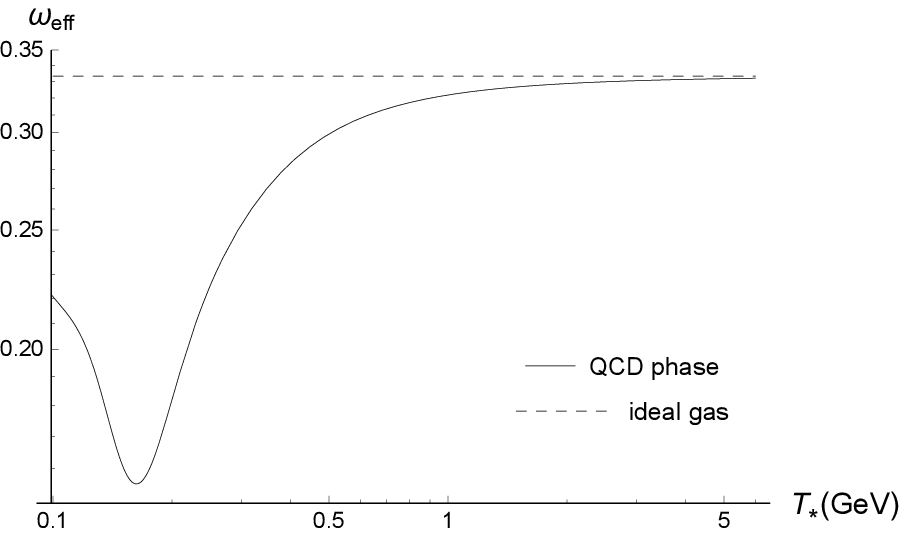}
    \caption{The effective equation of state parameter as a function of transition temperature $T_*$}
  \end{minipage}
  \hfill
  \begin{minipage}[b]{0.5\textwidth}
    \includegraphics[width=7cm, height=5cm]{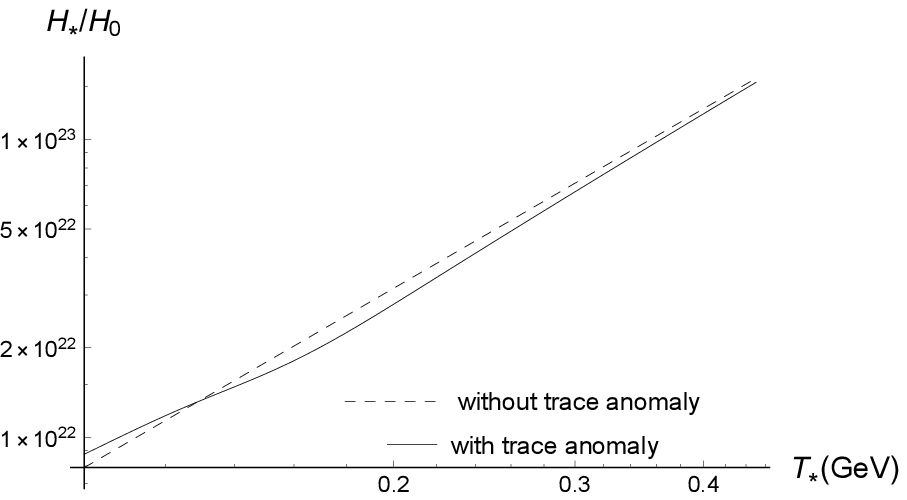}
    \caption{$\frac{H_*}{H_0}$ as a function of transition temperature $T_*$ without GUP effect $\beta=0$}
  \end{minipage}
\end{figure}
The peak frequency of the SGW frequency redshifted to the corresponding value at current epoch is given by
\begin{eqnarray}\label{b15}
\nonumber \frac{\nu_{0peak}}{\nu_*}=\frac{a_*}{a_0},~~~~~~~~~~~~~~~~~~~~~~~~~~~~~~~~~~~~~~~~\\
=\frac{T_0}{T_*}\left[\frac{g_s(T_0)}{g_s(T_*)}\right]^{\frac{1}{3}} \left[\frac{7+36\pi^2\beta T^2_0}{7+36\pi^2\beta T^2_*}\right]^{-\frac{1}{3}}.
\end{eqnarray}
 Fig. (3) shows the frequency received at the current time to that presents at the epoch of transition as a function of the transition temperature. It shows that the effect of GUP increases with increasing transition temperature. We have used the numerical values $T_0=2.725K=2.348\times 10^{-13}GeV$, $g_s(T_*)\in\left[33-37\right]\approx 35$ and $g_s(T_0)=3.4$.
According to Eq. (\ref{b3}), the function $\frac{a_0}{a_r}$ can be determined and used into Eq. (\ref{b11}) as
\begin{eqnarray}\label{b16}
\nonumber \left(\frac{H_*}{H_0}\right)^2=\Omega_{r0}\left(\frac{T_r}{T_0}\right)^4\left[\frac{g_s(T_r)}{g_s(T_0)}\right]^{\frac{4}{3}} \left[\frac{7+36\pi^2\beta T^2_r}{7+36\pi^2\beta T^2_0}\right]^{-\frac{4}{3}}\times ~~~~~~~~~~~~~~~~~~~~~~~~~~\\
\frac{g_s(T_*)^{1+\omega_{eff}(T_*)}}{g_s(T_r)^{1+\omega_{eff}(T_r)}}
\exp{\left[\int_{T_r}^{T_*}(1+\omega_{eff})\left(\frac{3}{T}-\frac{72\pi^2\beta T}{7+36\pi^2\beta T^2}\right)~dT\right]}.
\end{eqnarray}
Fig. (4) shows the ratio between Hubble parameter at the epoch of transition to its counterpart at current epoch as a function of transition time. It is clear that the effect of GUP increases with increasing transition temperature and leads to a decrease in the ratio with temperature. In other words, for high temperature, if $\beta_{1}>\beta_{2}>\beta_{3}$ it is concluded that $\frac{H_*}{H_0}|_{\beta_1}<\frac{H_*}{H_0}|_{\beta_2}<\frac{H_*}{H_0}|_{\beta_3}$.
Finally, we can  rewrite Eq. (\ref{b12}) as
\begin{eqnarray}\label{b17}
\nonumber \frac{\Omega_{gw}}{\Omega_{gw*}}=\Omega_{r0} ~
\left(\frac{T_r}{T_*}\right)^4\left[\frac{g_s(T_r)}{g_s(T_*)}\right]^{\frac{4}{3}} \left[\frac{7+36\pi^2\beta T^2_r}{7+36\pi^2\beta T^2_*}\right]^{-\frac{4}{3}}\times  ~~~~~~~~~~~~~~~~~~~~~~~~~~\\
\frac{g_s(T_*)^{1+\omega_{eff}(T_*)}}{g_s(T_r)^{1+\omega_{eff}(T_r)}}
\exp{\left[\int_{T_r}^{T_*}(1+\omega_{eff})\left(\frac{3}{T}-\frac{72\pi^2\beta T}{7+36\pi^2\beta T^2}\right)~dT\right]}.
\end{eqnarray}
\begin{figure}
\includegraphics[width=7cm, height=5cm]{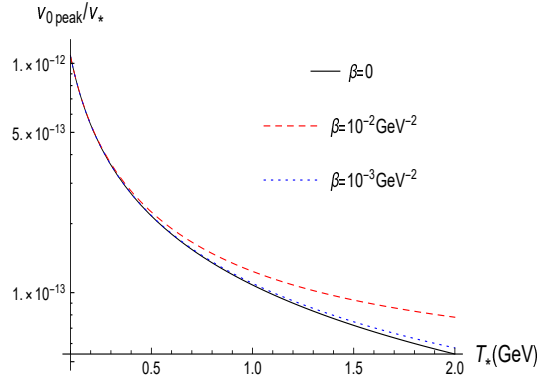}
\caption{$\frac{\nu_{0peak}}{\nu_*}=\frac{a_*}{a_0}$ as a function of transition temperature $T_*$ for some values of $\beta$}
\end{figure}
\begin{figure}[!tbp]
  \centering
  \begin{minipage}[b]{0.4\textwidth}
    \includegraphics[width=7cm, height=5cm]{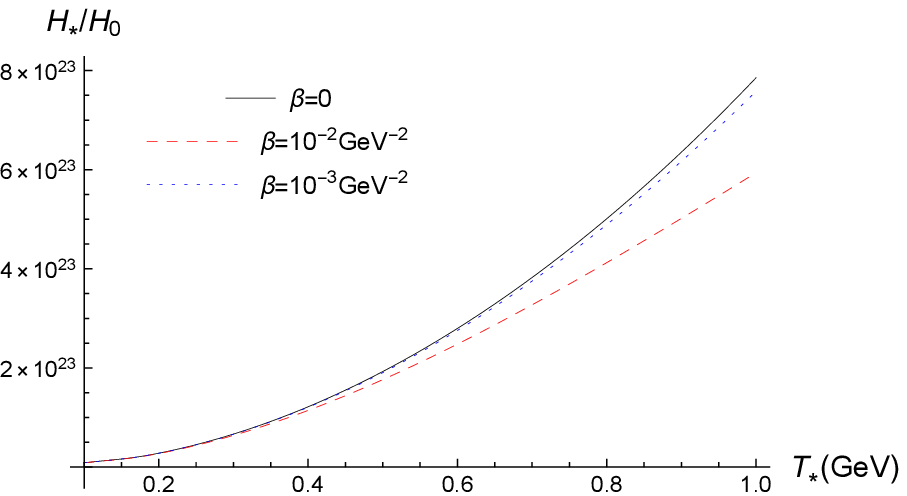}
    \caption{$\frac{H_*}{H_0}$ as a function of transition temperature $T_*$ using trace anomaly formalism with different values of $\beta$}
  \end{minipage}
  \hfill
  \begin{minipage}[b]{0.5\textwidth}
    \includegraphics[width=7cm, height=5cm]{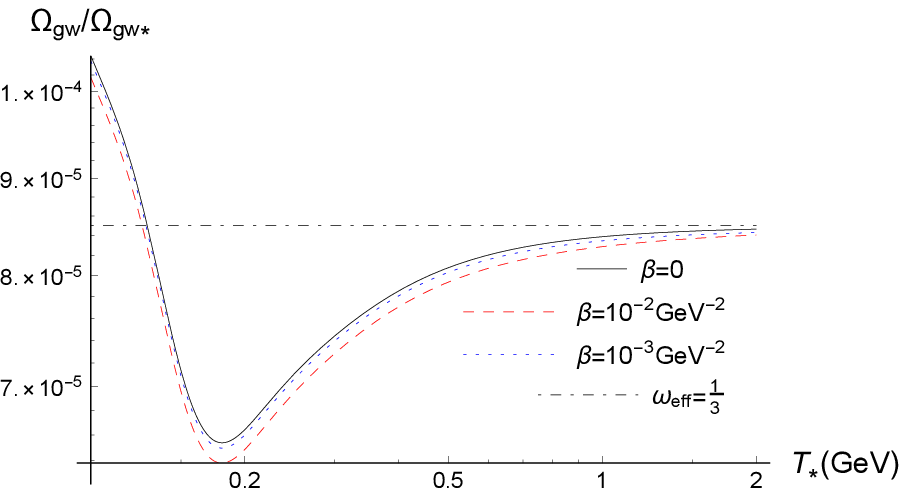}
    \caption{$\frac{\Omega_{gw}}{\Omega_{gw*}}$ as a function of transition temperature $T_*$ using trace anomaly formalism with different values of $\beta$ and the dotted dashed line represents the case $\omega_{eff}=\frac{1}{3}$ where $\frac{\Omega_{gw}}{\Omega_{gw*}}=\Omega_{r0}$}
  \end{minipage}
\end{figure}

Using trace anomaly formalism, the relative density parameter with different values of $\beta$ and applying equation of states of ultra-relativistic non-interacting gas is depicted in Fig. (5)
In the case of using equation of state for ultra-relativistic gas with non-interacting particles, $\omega_{eff}=\frac{1}{3}$, the two integrations in Eq. (\ref{b12}) will cancel each other and the relative density parameter will be constant, i.e. $\frac{\Omega_{gw}}{\Omega_{gw*}}=\Omega_{r0}=8.5\times 10^{-5}$. This shown in Fig. (5) as a straight line.
As the QCD equation of state goes to the equation of state of the ultra-relativistic and non-interacting gas, employing the QCD state equation, the relative density parameter also goes to the relative density parameter by using equation of state of ultra-relativistic and noninteracting gas, at $\sim5~GeV$. This behaviour is clearly shown in Fig. (5).
Also, we can notice that the relative Hubble parameter and density parameter ratio decrease due to the GUP effect. The reduction is always proportional to the GUP parameter $\beta$ as shown in Figs. (4) and (5). We have used the numerical values $T_r=10^4~GeV$ and $g_s(T_r)=106$.

\section{Modified QCD sources of stochastic gravitational wave}

In this section, we turn our attention to the density parameter of gravitational wave at the epoch of transition $\Omega_{gw*}=\Omega_{gw}(T_*)$, which is necessary to define the SGW spectrum through Eq. (\ref{b17}). There are many sources that contribute to the SGW background such as solitons and solitons stars \cite{c1}, cosmic strings and domain walls \cite{c2,c3}. Here we are interested in the SGW background that is generated due to the first order phase transition in the early universe. We consider two component sources of bubble percolation that take place after bubble nucleation and its expansion in the QCD first order phase transition, i.e, bubble wall collisions (BWC) and shocks in the plasma \cite{c4,c5,c6,c7,c8,c9}, and turbulent magnetohydrodynamics (MHD) in plasma after bubble collision \cite{c10}.

Based on envelope approximation and using  numerical simulation, the contribution to the SGW spectrum by bubble collisions is given by \cite{c9,c11,c13}
\begin{equation}\label{c1}
\Omega^{BWC}_{gw*}(\nu)=\left(\frac{H_*}{\gamma}\right)^2\left(\frac{\kappa_b\alpha}{1+\alpha}\right)^2
\left(\frac{0.11~v^3}{0.24+v^2}\right)S_{BWC}(\nu),
\end{equation}
where $\gamma^{-1}$ is the time duration of phase transition, $\kappa_b$ is the fraction of the latent heat of the phase transition deposited on the bubble wall, $\alpha$ is the ratio of the vacuum energy
density released in the phase transition to that of the radiation and $v$ is the wall velocity. Then, parameterization of the SGW spectrum is given by the function $S_{BWC}(\nu)$, which is determined by fitting simulation and analytically data, as
\begin{equation}\label{c2}
S_{BWC}(\nu)=\frac{3.8\left(\frac{\nu}{\nu_{BWC}}\right)^{2.8}}{1+2.8\left(\frac{\nu}{\nu_{BWC}}\right)^{3.8}},
\end{equation}
\begin{equation}\label{c3}
\nu_{BWC}=\frac{62~\gamma}{180-10v+100v^2}\left(\frac{a_*}{a_0}\right).
\end{equation}
It is worth noting that the frequency $\nu_{BWC}$ is the today peak frequency of the SGW that generated by BWC mechanism during phase transition. At the epoch of phase transition, the kinetic and magnetic Reynolds number of cosmic fluid are increased \cite{c14}, which motivate the bubbles to produce magnetohydrodynamical turbulence in the fully ionized plasm. Then assuming a presence of Kolmogorov-type turbulence, as suggested in \cite{c15}, the contribution to SGW spectrum form magnetohydrodynamical turbulence can be given as \cite{c14,c16}
\begin{equation}\label{c4}
\Omega^{MHD}_{gw*}(\nu)=\left(\frac{H_*}{\gamma}\right)
\left(\frac{\kappa_{MHD}\alpha}{1+\alpha}\right)^{\frac{3}{2}}v~S_{MHD}(\nu),
\end{equation}
where $\kappa_{MHD}$ is the fraction of latent heat that converted into turbulence, and the formula for the spectrum is given by
\begin{equation}\label{c5}
S_{MHD}(\nu)=\frac{\left(\frac{\nu}{\nu_{MHD}}\right)^3}{\left(1+\frac{\nu}{\nu_{MHD}}\right)^{\frac{11}{3}}\left[1+8\pi\frac{\nu}{H_*}\left(\frac{a_*}{a_0}\right)^{-1}\right]},
\end{equation}
\begin{equation}\label{c6}
\nu_{MHD}=\frac{7\gamma}{4v}\left(\frac{a_*}{a_0}\right),
\end{equation}
where $\nu_{MHD}$ is today's peak frequency of the SGW generated by MHD at the epoch of phase transition. Although parameters $\alpha$ and $\kappa$ play substantial roles in defining the peak position and amplitude of the SGW signal, there is still no surefire way to find $\kappa$. Following \cite{bc1}, we assume $v=0.7$, $\frac{\kappa_b\alpha}{1+\alpha}=\frac{\kappa_{MHD}\alpha}{1+\alpha}=0.05$, $\gamma=nH_*=5H_*$ and we use the following substitutions \cite{bc1,c17}
\begin{equation}\label{c7}
H_*=\sqrt{\frac{8\pi}{3m_p^2}\rho(T_*)},~~~~T_*=T_c,
\end{equation}
where
\begin{equation}\label{c8}
\rho(T_*)=T_*^5\frac{dF(T_*)}{dT_*}+3T_*^4F(T_*).
\end{equation}
Now, using the above equations, we obtain
\begin{equation}\label{c9}
\Omega^{BWC}_{gw*}(\nu)=10^{-4}
\frac{0.11~v^3}{0.24+v^2}~
\frac{3.8\left(\frac{\nu}{\nu_{BWC}}\right)^{2.8}}{1+2.8\left(\frac{\nu}{\nu_{BWC}}\right)^{3.8}},
\end{equation}
\begin{equation}\label{c10}
\Omega^{MHD}_{gw*}(\nu)=\frac{0.05^{\frac{3}{2}}}{5}
\frac{v\left(\frac{\nu}{\nu_{MHD}}\right)^3}{\left(1+\frac{\nu}{\nu_{MHD}}\right)^{\frac{11}{3}}
\left[1+\frac{8\pi\nu}{5}\sqrt{\frac{3m_p^2}{8\pi}\frac{1}{\rho(T_*)}}\left(\frac{a_*}{a_0}\right)^{-1}\right]},
\end{equation}
where
\begin{equation}\label{c11}
\nu_{BWC}=\frac{310}{180-10v+100v^2}\sqrt{\frac{8\pi}{3m_p^2}\rho(T_*)}\left(\frac{a_*}{a_0}\right),
\end{equation}
\begin{equation}\label{c12}
\nu_{MHD}=\frac{35}{4v}\sqrt{\frac{8\pi}{3m_p^2}\rho(T_*)}\left(\frac{a_*}{a_0}\right).
\end{equation}
Then, we use $\Omega_{gw*}h^2=(\Omega^{BWC}_{gw*}(\nu)+\Omega^{MHD}_{gw*}(\nu))h^2$ into Eq. (\ref{b17}) and take into account Eq. (\ref{b3}). Using the above mentioned numerical values of the parameters, we estimate the SGW spectrum $\Omega_{gw}h^2$, as shown in Fig. (6). It is clear that using QCD equation of state with trace anomaly causes the SGW signal becomes weak with increasing frequency. This effect is enhanced by the presence of quantum gravity effect, where the SGW signal weakens more compared to normal case. Using Eqs. (\ref{c11}) and (\ref{c12}) we obtain the total peak frequency of the SGW signal that can be measured today generated from BWC and MHD mechanism at the epoch of phase transition as
\begin{eqnarray}
\nonumber \nu_{total}= \nu_{BWC}+\nu_{MHD},~~~~~~~~~~~~~~~~~~~~~~~~~~~~~~~~~~~~~~~~~~~~~~~~~~~~~~~~~~~~~~~~~~~~~~~~~~~~~~~~~~~\\
=\left(\frac{310}{180-10v+100v^2}+\frac{35}{4v}\right)
\frac{T_0}{T_*}\left[\frac{g_s(T_0)}{g_s(T_*)}\right]^{\frac{1}{3}} \left[\frac{7+36\pi^2\beta T^2_0}{7+36\pi^2\beta T^2_*}\right]^{-\frac{1}{3}}
\sqrt{\frac{8\pi}{3m_p^2}\rho(T_*)}.
\end{eqnarray}
Fig. (7) shows the influence of GUP on the peak signal of SGW that can be measured today as a function of the transition temperature.
This effect becomes significant as the transfer temperature increases or as the GUP parameter increasing. In other words, at high limit $T_*$, if $\beta_{1}<\beta_{2}$ it is concluded that for fixed temperatures, $\nu_{total}(\beta_{1})<\nu_{total}(\beta_{2})$. Also, at this limit, the effects of GUP increases the peak of frequency i.e., $\nu_{total}^{\beta=0}<\nu_{total}^{\beta\neq0}$. These differences become dominant with increasing temperature.

 It is useful to compare the our result of $\Omega_{gw*}h^2$ in the presence of GUP effect with the possible sensitivities of interferometers designed to probe the stochastic background, such as  LIGO-VIRGO or  LISA satellites. It is found in \cite{Capozziello:2017vdi} the sensitivity is bounded by the value $10^{-12}$ as set by LIGO-VIRGO. It is also found that old LISA configuration over  a year can  can detect a white-noise stochastic background at the level of $10 ^{-13}$ \cite{Caprini:2015zlo}. It is clear that these sensitives values would set a stringent abound on the GUP parameter $\beta$ which may signify an intermediate length scale between Planck scale and electroweak scale which  may be consistent with that set by the electroweak scale. The stochastic background sensitivity in that sense may have interesting implications with minimal length theories.  We hope to study this in details in the near future.

\begin{figure}[!tbp]
  \centering
  \begin{minipage}[b]{0.4\textwidth}
    \includegraphics[width=7cm, height=5cm]{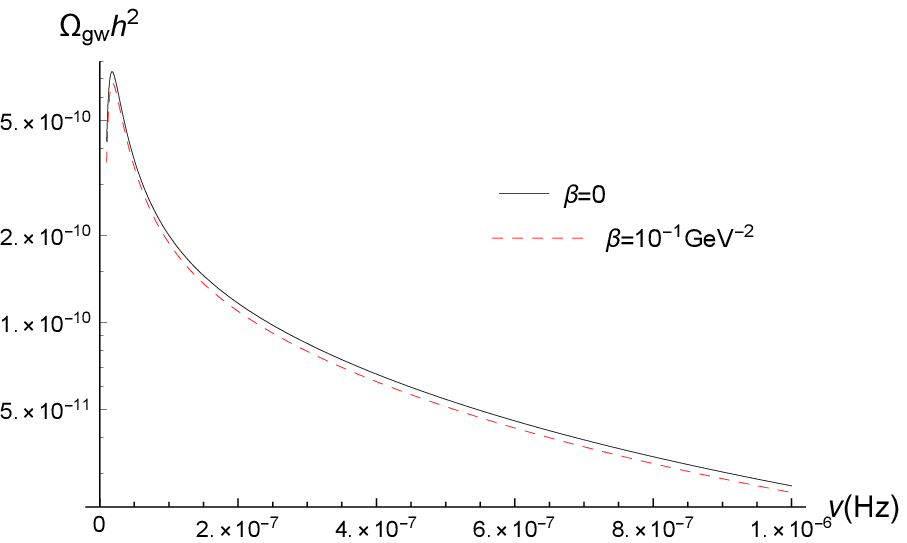}
    \caption{Net  contribution to the SGW due to bubble wall collision and
magnetohydrodynamic turbulence}
  \end{minipage}
  \hfill
  \begin{minipage}[b]{0.5\textwidth}
    \includegraphics[width=7cm, height=5cm]{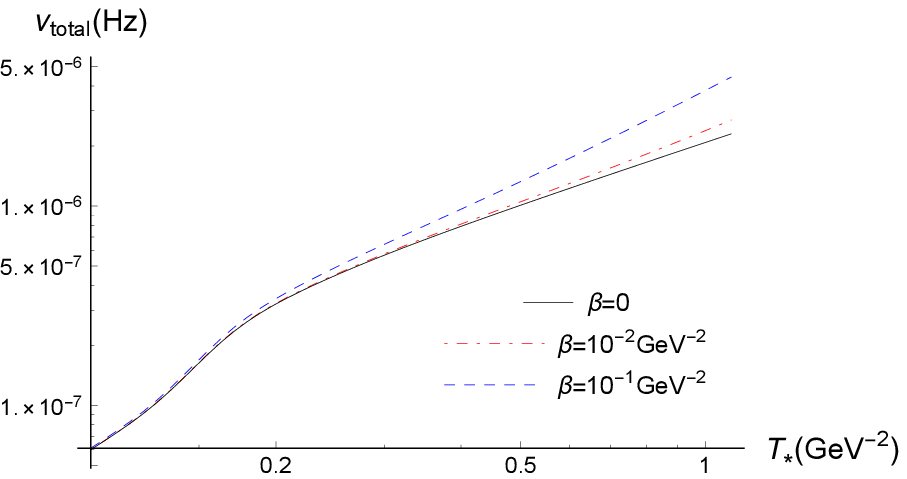}
    \caption{Today's net peak frequency of SGW signal arising from BWC and MHD at the epoch of phase transition}
  \end{minipage}
\end{figure}

\section{Conclusion}

In this paper, we have studied the effects of GUP on the SGW background signal generated at the quark-hadron phase transition, which corresponds to the cosmological first order phase transition. It was agreed that such a cosmological phase transition occurred at the early universe within QCD scale about $t\sim 10^{-5}s$ after the big bang at temperature $T\sim 0.2~GeV$ and the Hubble radius in order $10\sim Km$, which involving a mass about $M_{\odot}$.

Within GUP framework, the modified thermodynamical properties of photons gas is investigated.
Using the resulting modified entropy, the temporal evolution of the universe temperature calculated as a function of the Hubble parameter.
Due to the effects of QCD interaction, at epoch of phase transition, QCD equation of state should be employed. A pressure due to strong interactions among  massless $u,~d,~s$ quarks and gluons, with energy density resulting from trace anomaly, was used to formulate the QCD equation of state.
Within GUP framework and QCD equation of state, evolution of Hubble parameter and the energy density parameter of stochastic gravitational wave from the epoch of phase transition till current time was investigated.
It is found that the presence of GUP leads to decrease in relative Hubble parameter and energy density parameter ratio, and the reduction is always proportional to the value of GUP parameter.
A redshift in the peak frequency of the SGW at current epoch was obtained. Results showed that GUP effects caused an increase in the frequency peak which can lead to the better detection of the SGW signal.
There are a various mechanism that can produce stochastic gravitational wave at epoch of transition. In this paper we were interested in BWC and MHD as a sources of the gravitational waves. Within GUP framework, the modified expressions of BWC and MHD, that contribute in the SGW spectrum, were calculated. It is found that the SGW signal generated during QCD phase transition became weaker in comparison to its counterpart for $\beta=0$. Today's net peak frequency of SGW signal, which was produced from BWC and MHD at the epoch of phase transition, is investigated. In the presence of GUP, it is found that the frequency of SGW signal is increased in comparison with original case, and the growth in the frequency depends on the GUP parameter.
We also found that the sensitivities of interferometers designed to probe the stochastic background could set a stringent bound on the GUP parameter. These results could shed light on increasing the chance of detecting the stochastic gravitational signal created by such a transition in future observations. Moreover, they can lead to decoding of the dynamics of QCD phase transition at the early universe.

\end{document}